\documentclass[letterpaper,onecolumn, 16 pt, conference]{article}
\usepackage{graphicx}
\usepackage{epstopdf}
\usepackage{multicol}
\usepackage{stfloats}
\usepackage{amsfonts}
\usepackage{mathrsfs}
\usepackage{multicol}
\usepackage{stfloats}
\usepackage{cite}
\usepackage{amsfonts}
\usepackage{amsthm}
\usepackage{mathrsfs}
\usepackage{amsfonts}
\usepackage{empheq}
\usepackage{color}
\usepackage[ruled]{algorithm2e}
\usepackage{fancybox}
\usepackage{framed}
\usepackage{amsfonts, color}
\usepackage{color}
\usepackage{setspace}
\usepackage[dvips]{epsfig}
\usepackage{psfrag, amssymb, amsmath, cite}
\usepackage[colorlinks,linkcolor=red,anchorcolor=blue,citecolor=blue]{hyperref}
\usepackage{verbatim}

\usepackage[margin=0.75in]{geometry}

\newtheorem{remark}{Remark}
\newtheorem{definition}{\textbf{Definition}}

\newtheorem{fact}{\textbf{Fact}}
\newtheorem{example}{\textbf{Example}}

\title{\bf Identification of Hessian matrix in distributed gradient-based multi-agent coordination control systems}

\author{\textbf{Zhiyong Sun} and \textbf{Toshiharu Sugie} 
\thanks{Zhiyong Sun is with Research School of Engineering, The Australian
National University, Canberra ACT 2601, Australia. Email: {\tt\small zhiyong.sun@anu.edu.au, sun.zhiyong.cn@gmail.com}. }  
\thanks{
Toshiharu Sugie is with Graduate School of Informatics, Kyoto University, Yoshida-honmachi, Sakyo-ku, Kyoto 606-8501 Japan. Email: {\tt\small sugie@i.kyoto-u.ac.jp}.  } }

\begin{document}
\date{}
\maketitle

\begin{abstract}
Multi-agent coordination control usually involves a potential function that encodes information of a global control task, while the control input for individual agents is often designed by a gradient-based control law. The property of Hessian matrix associated with a potential function  plays an important role in the stability analysis of equilibrium points in gradient-based coordination control systems. Therefore, the identification of Hessian matrix in gradient-based multi-agent coordination systems becomes a key step in multi-agent equilibrium analysis. However, very often the identification of Hessian matrix via the entry-wise calculation is  a very tedious task and can easily introduce calculation errors.  In this paper we present some general and fast approaches for the identification of Hessian matrix based on matrix differentials and calculus rules, which  can  easily derive a compact form of Hessian matrix for multi-agent coordination systems. We also present several examples on Hessian identification for certain typical potential functions involving edge-tension distance functions and triangular-area functions, and  illustrate their applications   in the context of distributed coordination and formation control. 
\end{abstract}

\section{Introduction}
\subsection{Background and related literature}
In recent years cooperative coordination and distributed control for  networked multiple agents (e.g., autonomous vehicles or mobile robots etc.) have gained considerable attention in the control, optimization and robotics community \cite{cao2013overview, knorn2016overview}. This has been motivated by various applications such as formation control, coordination in complex networks, sensor networks, distributed optimization, etc. A typical approach for designing distributed control law for coordinating individual agents is to associate an  objective potential function for the whole multi-agent  group, while the control law for each individual agent is a gradient-descent law that minimizes the specified potential function \cite{sugie2015gradient, Xudong_gradient}. Very often, such potential functions are defined by geometric quantities such as distances or areas related with agents' positions over an interaction graph in the configuration space. Typical scenarios involving gradient-based control in multi-agent coordination include distance-based formation control \cite{krick2009stabilisation, Sun2016exponential, sun2017distributed, anderson2017formation, Xudong_SIAM_triangle}, multi-robotic maneuvering and manipulability control  \cite{KAWASHIMA2014}, motion coordination with constraints \cite{Zhao2017general}, among others. Comprehensive discussions and solutions  to characterize distributed gradient control laws for multi-agent coordination control are provided in\cite{sugie2015gradient} and   \cite{sugie2018TCNS}, which  emphasize the notion of clique graph (i.e., complete subgraph) in designing potential functions and gradient-based controls.  The recent book \cite{sun2018cooperative} provides an updated review on recent progress of cooperative coordination and distributed control of multi-agent systems. 

For multi-agent coordination control in a networked environment, a key task in the control law design and system dynamics  analysis is to determine convergence and stability of such gradient-based multi-agent systems with a group potential function. Gradient systems enjoy several nice convergence properties and can guarantee local convergence if certain properties such as positivity and analyticity of potential functions are satisfied. However, in order to determine stability of different equilibrium points  of gradient systems, Hessian matrix of potential functions are necessary and should be identified. 

For gradient systems, Hessian matrix plays an important role in determining whether an equilibrium point is stable or unstable (i.e., being a saddle point etc). Hessian also provides key information to reveal more properties (such as hyperbolicity) of an equilibrium associated with a potential function.  However, identification of Hessian matrix is  a non-trivial and often  very tedious task, which becomes even more involved in the context of multi-agent coordination control, in that graph topology that models agents' interactions in a networked manner should also be taken into consideration in the Hessian formula.
The standard way of Hessian identification usually involves entry-wise calculation, which we refer as `direct' approach. But this approach soon becomes intractable when a multi-agent coordination system under consideration involves complicated dynamics, and the interaction graph   grows in size with more complex topologies. Alternatively, matrix calculus that takes into account graph topology and coordination laws can offer a more convenient approach in identifying Hessian matrices and deriving a compact Hessian formula, and this motivates this paper. 

In this paper, with the help of matrix differentials and calculus rules, we discuss Hessian identification for several typical potentials commonly-used in gradient-based multi-agent coordination control. 
We do not aim to provide a comprehensive study on Hessian identification for multi-agent coordination systems, but we will identify Hessian matrices for two general potentials associated with an underlying undirected graph topology. The first is an edge-based, distance-constrained potential that is defined by an edge function for a pair of agents, usually involving inter-agent distances. The overall potential is a sum of all individual potentials over all edges.  The second type of potential function is defined by a three-agent subgraph, usually involving the (signed) area quantity spanned by a three-agent subgraph. We will use the formation control with signed area constraints as an example of such distributed coordination systems, and illustrate how to derive Hessian matrix for these coordination potentials in a general graph. The identification process of Hessian formula   can be extended in  identifying other Hessians matrices in even more general potential functions used in multi-agent coordination control.



\subsection{Paper contributions and organizations}
The main contributions of this paper include the following. 
We will first present two motivating examples with comparisons on different identification approaches, in which we favor the `indirect' approach based on matrix calculus in the identification.
For some typical multi-agent potentials defined as edge-tension, distance-based functions, we will   derive a general formula of Hessian matrix that can be readily applied in calculating Hessians for   potential functions with particular terms. For potential functions involving both distance functions and triangular-area functions, we will show, by using two representative examples, how a compact form of Hessian matrix can be obtained by following basic matrix calculus rules. 
Note it is not the aim of this paper to cover all different types of potentials in multi-agent coordination and identify their Hessian formulas. Rather, apart from the identification results of several Hessians, the paper will also serve as a tutorial on Hessian identification for multi-agent coordination systems by analyzing some representative potential functions, and by following matrix calculus rules we will aim to advance this approach in Hessian identification in the context of multi-agent coordination control.

This paper is organized as follows. Section \ref{sec:background_matrix} reviews several essential tools of matrix/vector differentials and calculus rules that will be used in the derivation of Hessian matrix for various potential functions.  Section \ref{sec:preliminary_graph_gradient} presents 
preliminaries on basic graph theoretic tools in modeling multi-agent distributed systems, and gradient systems for designing gradient-distributed controllers for multi-agent coordination control. Motivating examples with a two-agent system and with a three-agent system are discussed in Section \ref{sec:motivating_examples}, which presents obvious advantages of using matrix calculus rules in identifying Hessian matrix for multi-agent coordination potentials. Section \ref{sec:Hessian_edge} discusses a unified and general formula of Hessian identification for edge-tension, distance-based potentials that are commonly-used in modeling multi-agent coordination tasks. Several typical examples of edge-based potentials are also discussed in this section, with their Hessian matrices correctly identified by following the derived general formula. Section \ref{sec:Hessian_composite} shows general approaches for identifying Hessian matrix for composite potential functions that involve not only edge-based distance  functions but also triangular-area-based functions within three-agent groups as complete subgraphs. Brief discussions and   remarks are shown in  \ref{sec:conclusions} that conclude this paper.

\subsection{Notations}
The notations used in this paper are fairly standard. A real scalar valued function $f$ is called a $C^r$ function if it has continuous first $r$ derivatives. The notation `d' denotes `differential'. 
We use $\mathbb{R}^n$ to denote the $n$-dimensional Euclidean space, and $\mathbb{R}^{m\times n}$
to denote the set of $m\times n$ real matrices. The transpose of a  matrix or vector $M$ is denoted by  $M^{\top}$.
For a vector $v$, the symbol $\|v\|$ denotes its  Euclidean norm.
  We denote the $n \times n$ identity matrix  as  $I_{n}$.  
 A diagonal matrix obtained from an $n$-tuple  vector $\{x_1, x_2, \cdots, x_n\}$ with $x_i \in \mathbb{R}$ as its diagonal entries is denoted as $\text{diag}(x_1, x_2, \cdots, x_k) \in \mathbb{R}^{n \times n}$, and a \textit{block} diagonal matrix obtained from $n$-column $d$-dimensional vectors $\{x_1, x_2, \cdots, x_n\}$ with $x_i \in \mathbb{R}^d$  as its diagonal \textit{block} entries is denoted as $\text{blk-diag}(x_1, x_2, \cdots, x_k) \in \mathbb{R}^{dn \times n}$. 
The symbol $\otimes$  denotes   Kronecker product. 

\section{Background on vector/matrix differentials} \label{sec:background_matrix}
In this section we review some background on matrix calculus, in particular some fundamental rules on  vector/matrix differentials. More discussions and properties on matrix calculus can be found in \cite[Chapter 3]{zhang2017matrix}, \cite[Chapter 15]{harville1998matrix}, and \cite[Chapter 13]{matrix_2005}.  

Consider a real scalar function $f(x): \mathbb{R}^m \rightarrow \mathbb{R}$ that is differentiable with the variable $x=[x_1, \ldots,x_m]^\top \in \mathbb{R}^m$. \footnote{One sufficient condition for a multivariate function $f(x_1,\ldots,x_m)$ to be differentiable at the point $(x_1,\ldots,x_m)$ is that the partial derivatives $\partial f/\partial x_1,\ldots,\partial f/\partial x_m$ exist and are continuous.}  The first-order differential (or simply differential) of the multivariate function $f(x_1,\ldots,x_m)$   is denoted by
\begin{align}
\text{d} f(x) = \frac{\partial f(x)}{\partial x_1} \text{d} x_1 + \cdots + \frac{\partial f(x)}{\partial x_m} \text{d} x_m = \left[\frac{\partial f(x)}{\partial x_1},  \cdots, \frac{\partial f(x)}{\partial x_m} \right]  \left[
\begin{array}{c}
\text{d} x_1 \\
\vdots \\
\text{d} x_m
\end{array}
\right], 
\end{align}
or in a compact form
\begin{align} \label{eq:jacobian_definition}
\text{d} f(x) =  \frac{\partial f(x)}{\partial x^\top} \text{d} x  = (\text{d} x)^\top \frac{\partial f(x)}{\partial x},
\end{align}
where $\frac{\partial f(x)}{\partial x^\top} :=\left[\frac{\partial f(x)}{\partial x_1},  \cdots, \frac{\partial f(x)}{\partial x_m} \right]$  and $\text{d} x := [\text{d} x_1, \cdots, \text{d} x_m]^\top$. In this way one can identify the Jacobian matrix $D_x f(x) := \frac{\partial f(x)}{\partial x^\top} \in \mathbb{R}^{1 \times m}$, which is a \textit{row} vector. 
According to convention, we also denote the gradient vector as a \textit{column} vector, in the form $\nabla_x f(x): =  \left[\frac{\partial f(x)}{\partial x_1},  \cdots, \frac{\partial f(x)}{\partial x_m} \right]^\top \in \mathbb{R}^{m \times 1}$. 

Note the same rule can also be applied to the identification of Jacobian matrix for a real vector-valued function $f(x): \mathbb{R}^m \rightarrow \mathbb{R}^n$, in which  the Jacobian matrix can be identified as 
 $D_x f(x) := \frac{\partial f(x)}{\partial x^\top} \in \mathbb{R}^{n \times m}$.
 
Now we consider a real scalar function $f(x) \in C^2: \mathbb{R}^m \rightarrow \mathbb{R} $ (i.e., twice differentiable functions). We denote  the Hessian matrix, i.e., the second-order derivative of a real function
$f(x)$, as $\mathcal{H}_{f(x)}$, which is defined as
\begin{align}
\mathcal{H}_{f(x)} = \frac{\partial^2 f(x)}{\partial x \partial x^\top} = \frac{\partial}{\partial x} \left(\frac{\partial f(x)}{\partial x^\top}\right) \in \mathbb{R}^{m \times m}.
\end{align}
In a compact form, we can also write
\begin{align}
\mathcal{H}_{f(x)} = \nabla^2_x f(x) = \nabla_x(D_x f(x)).
\end{align}
Therefore, the $(i,j)$-th entry of $\mathcal{H}$ is defined as
\begin{align} \label{eq:hessian_definition}
\mathcal{H}_{f(x),ij} & = \left[\frac{\partial^2 f(x)}{\partial x \partial x^\top}\right]_{ij} = \frac{\partial}{\partial x_i}\left(\frac{\partial f(x)}{\partial x_j}\right) \nonumber \\
& = \left[\frac{\partial^2 f(x)}{\partial x \partial x^\top}\right]_{ji} = \frac{\partial}{\partial x_j}\left(\frac{\partial f(x)}{\partial x_i}\right), 
\end{align}
where the equality in the second line is due to the symmetry of Hessian matrix. 

The entry-wise definition of Hessian $\mathcal{H}_f$ in \eqref{eq:hessian_definition}  presents a standard and direct approach to identify the Hessian matrix   for a real scalar function $f$. However, in general it is not convenient for performing the calculation in practice by following the entry-wise definition  \eqref{eq:hessian_definition}. We will now discuss a faster and more efficient approach for Hessian matrix identification based on matrix calculus rules. 

From the compact form of first-order differential $\text{d} f(x)$ in \eqref{eq:jacobian_definition}, one can calculate the second-order differential as 
\begin{align}  \label{eq:hessian_basic}
\text{d}^2 f(x) = \text{d}(\text{d} f(x)) &= \underbrace{\text{d} (\text{d} x^\top)}_{=0} \frac{\partial  f(x)}{\partial x} + \text{d} x^\top \frac{\partial \text{d} f(x)}{\partial x} \nonumber \\
&= (\text{d} x)^\top \frac{\partial }{\partial x} \left(\frac{\partial f(x)}{\partial x^\top} \right) \text{d} x \nonumber \\
&= (\text{d} x)^\top  \underbrace{  \frac{\partial^2 f(x)}{\partial x \partial x^\top}}_{ :=\mathcal{H}_{f}}   \text{d} x  
\end{align}
which presents a quick and convenient way to identify Hessian matrix in a compact form.  Note that in the above derivation we have used the fact $\text{d} (\text{d} x^\top) = 0$ because $\text{d} x$ is not a function of the vector $x$. 	In this paper, we will frequently use  \eqref{eq:hessian_basic} to identify Hessian matrices for several typical potential functions applied in multi-agent coordination control. 

\section{Preliminaries on graph theory and gradient systems}  \label{sec:preliminary_graph_gradient}
\subsection{Basic graph theoretic tools and applications in modeling multi-agent systems}
Interactions in multi-agent coordination systems are usually modeled by graphs, for which  we review several graph theoretic tools in this section. Consider an undirected graph with $m$ edges and $n$ vertices, denoted by $\mathcal{G} =( \mathcal{V}, \mathcal{E})$  with vertex set $\mathcal{V} = \{1,2,\cdots, n\}$ and edge set $\mathcal{E} \subset \mathcal{V} \times \mathcal{V}$. Each vertex represents an agent, and the edge set represents communication or interaction relationship between different agents. The neighbor set $\mathcal{N}_i$ of vertex $i$ is defined as $\mathcal{N}_i: = \{j \in \mathcal{V}: (i,j) \in \mathcal{E}\}$.
  The matrix relating the vertices to the edges is called the incidence matrix $H = \{h_{ij}\} \in \mathbb{R}^{m \times n}$, whose entries are defined as (with arbitrary edge orientations)
     \begin{equation}
     h_{ij} =  \left\{
       \begin{array}{cc}
       1,  &\text{ the } i\text{-th edge sinks at vertex }j;  \\  
       -1,  &\text{ the } i\text{-th edge leaves  vertex }j; \\  
       0,  & \text{otherwise}.  \\
       \end{array}
      \right.
      \end{equation}
Another important matrix representation of a graph $\mathcal{G}$ is the Laplacian matrix $L(\mathcal{G})$ \cite{mesbahi2010graph}. For an undirected graph, the associated Laplacian matrix can be written  as $L(\mathcal{G}) = H^{\top}H$.
For more introductions on  algebraic graph theory and their applications in distributed multi-agent systems and networked coordination control, we refer the readers to \cite{mesbahi2010graph} and \cite{bapat2010graphs}.

Let $p_i  \in \mathbb{R}^d$  denote a point  that is assigned to agent $i \in \mathcal{V}$ in  the $d$-dimensional Euclidean space $\mathbb{R}^d$.     The stacked  vector $p=[p_1^{\top}, \,
p_2^{\top}, \cdots, \, p_n^{\top}]^{\top} \in \mathbb{R}^{dn}$ represents a configuration of $\mathcal{G}$  realized in $\mathbb{R}^d$. 
Following the definition of the  matrix $H$, one can construct the relative position vector as an image of $H\otimes I_d$ from the position vector $p$:
\begin{equation}
z = (H\otimes I_d) p,  \label{z_equation}
\end{equation}
where $z=[z_1^{\top}, \,
z_2^{\top}, \cdots, \, z_m^{\top}]^{\top} \in \mathbb{R}^{dm}$, with $z_k  \in \mathbb{R}^d$ being the relative position vector for the vertex pair $(i,j)$ defined for the $k$-th edge: $z_k = p_i - p_j$. In this paper we may also use notations such as $z_{k_{ij}}$ or $z_{ij}$ if no confusion arises.

\subsection{Gradient systems and gradient-based multi-agent coordination control}
In this section we briefly review the definition and properties of gradient systems. Let $V(x): \mathbb{R}^n \rightarrow \mathbb{R}_{\geq 0}$ be   a scalar valued function   that is $C^r$ with $r \geq 2$.   Consider the following continuous-time system
\begin{align} \label{eq:gradient_model}
\dot x = -\nabla_x V(x).
\end{align}
The above system is usually called a \emph{gradient system}, and the corresponding function $V(x)$ is   referred to as a \emph{potential function}.

Gradient system enjoys several  convergence properties due to the special structure of the gradient vector field in the right-hand side of \eqref{eq:gradient_model}. Firstly, it should be clear that  equilibrium points of \eqref{eq:gradient_model} are critical points of $V(x)$. Moreover, at any point except for an equilibrium point,
the vector field \eqref{eq:gradient_model} is perpendicular to the level sets of $V(x)$. In fact, it is obvious to observe that $\dot V(x) = \nabla_x V(x)^{\top} \dot x = - \|\nabla_x V(x)\|^2 \leq 0$, which indicates that the potential $V(x)$ is always non-increasing along the trajectory of \eqref{eq:gradient_model}. The following results are also obvious.

\begin{fact}
Consider the gradient system \eqref{eq:gradient_model} with the associated potential $V(x)$.
\begin{itemize}
\item $\dot V(x) \leq 0 $ and $\dot V(x) = 0$ if and only if $x$ is an equilibrium point of \eqref{eq:gradient_model}.
\item Suppose $\bar x$ is an \emph{isolated minimum} of a real analytic $V (x)$, i.e., there
is a neighborhood of $\bar x$ that contains no other minima of $V (x)$. Then $\bar x$ is
an asymptotically stable equilibrium point of \eqref{eq:gradient_model}.
\end{itemize}
\end{fact}
The proof of the above facts can be found in e.g. \cite[Chapter 15]{wiggins2003introduction}.
Note that in the second statement we have emphasized the condition \emph{isolated minimum} in the convergence property. 
We also refer the readers to the book \cite[Chapter 15]{wiggins2003introduction} for more introductions and properties on gradient vector fields and gradient systems. 

Note that a  local minimum of $V$   is not necessarily a stable equilibrium point of \eqref{eq:gradient_model}, unless some more properties on the potential $V$ are imposed (while the smoothness of the potential $V$ is not enough).   In \cite{absil2006stable}, several  examples (and counterexamples) are carefully constructed to show the relationship between local minima of $V$ and stable equilibrium points of \eqref{eq:gradient_model}. In particular, it is shown in \cite{absil2006stable} that with  the analyticity \footnote{A real  function  is analytic if  it  possesses  derivatives  of  all  orders and agrees with its Taylor series in the neighborhood of every  point in its domain.} of the potential $V$,  local minimality becomes a necessary and sufficient condition for stability.
\begin{fact} \label{thm:analyticity_minimality} (see \cite[Theorem 3]{absil2006stable})
Let $V$ be real analytic in a neighborhood of an equilibrium $\bar x \in \mathbb{R}^n$. Then, $\bar x$ is a stable equilibrium point of \eqref{eq:gradient_model} if and only if it is a local minimum of $V$.
\end{fact}
In order to determine convergence and stability properties for general equilibrium points for a gradient system \eqref{eq:gradient_model}, one needs to further analyze the linearization matrix of \eqref{eq:gradient_model} (i.e., the Hessian matrix of $V$, with a reverse sign). Therefore, identification of Hessian matrix is a key step prior to analyzing equilibrium and convergence properties of gradient systems. 

In the context of multi-agent coordination control, gradient systems and gradient-based control provide a natural solution to coordination controller design. Very often,  group objective functions for a multi-agent system serve as a potential function, and control input for each agent typically involves a gradient-descent control that aims to minimize a specified potential function. A key question is whether the gradient control input for each agent is local and distributed, in the sense that control input only involves information (or relative information) of an agent itself and its neighbors as described by the underlying network graph that models interactions between individual agents. This question is addressed in \cite{sugie2015gradient}, from which we recall some key definitions and results as follows.
 The following definition refers to a fundamental property of objective potential functions whose gradient-based controllers \eqref{eq:gradient_model} are distributed.
 
\begin{definition} 
A class $C^1$ function $f_i$ is called gradient-distributed over the graph $\mathcal{G}$ if and only if its gradient-based controllers \eqref{eq:gradient_model} are distributed;  that is, there exist $n$ functions $f_i$ such that
\begin{align} \label{eq:gradient_distributed_definition}
\frac{\partial V(p)} {\partial p_{i}}=-f_{i}(p_{i},p_{{\cal N}_{i}}), \forall i\in {\cal N}. 
\end{align}
\end{definition} 
The recent papers \cite{sugie2015gradient} and \cite{sugie2018TCNS} provide a comprehensive study on gradient-based distributed control, in which   
a full characterization of the class of all gradient-distributed objective potential functions is discussed. A key result in \cite{sugie2015gradient} and \cite{sugie2018TCNS} is that   the notion of \textit{clique} (i.e., complete subgraph) plays a crucial role to obtain a distributed controller for multi-agent coordination control. That is, in order for a gradient-based coordination control to be distributed, the objective potential function should involve only agents' states in a clique. Typical cliques include edges associated with two agents, triangular subgraphs associated with three agents, etc. In this paper, our focus will be on the Hessian analysis of a distributed gradient-based coordination  control system \eqref{eq:gradient_distributed_definition} associated with an overall potential function, with the aim of providing some unified formulas of Hessian matrix. The identification of Hessian formulas will aid the stability analysis of different equilibriums in gradient-distributed multi-agent systems. 

\section{Motivating examples: Hessian matrix identification for simple gradient-based coordination systems}  \label{sec:motivating_examples}
\subsection{Hessian identification for a two-agent coordination system} \label{sec:motivating_examples_2agent}
As a motivating example,   we provide a general approach to identify   Hessians for simple gradient-based control systems that involve two or three agents (examples taken from \cite{Sugie2018CDC}). 
Consider a multi-agent system that consists of two agents $i$
and $j$ in a 2-D space, with $p_i \in \mathbb{R}^2$ being fixed and $p_j \in \mathbb{R}^2$  governed by
\begin{align} \label{eq:gradient_example_2agent1}
\dot p_j = -\nabla_{p_i} V_{ij} = -\frac{\partial V_{ij}}{\partial p_j},
\end{align}
where 
\begin{align} \label{eq:potential_sugie1}
V_{ij} = \frac{1}{4}\left(\|p_i - p_j\|^2 - d_{ij}^2 \right)^2. 
\end{align}
in which $d_{ij}$ is a positive value denoting a desired distance between agents $i$ and $j$. 

The gradient vector is
\begin{align}
\nabla_{p_i} V_{ij} = (\|p_i - p_j\|^2 - d_{ij}^2) (p_i - p_j).
\end{align}
Now we identify the Hessian matrix by following the matrix calculus rule in \eqref{eq:hessian_basic}:
\begin{align} \label{eq:hessian_example1}
\text{d}^2 V_{ij} &= (\text{d} p_i)^\top  \text{d} \nabla_{p_i} V_{ij} \nonumber \\
&= (\text{d} p_i)^\top \left(\text{d} \left(\|p_i - p_j\|^2 - d_{ij}^2 \right) (p_i - p_j) + \left(\|p_i - p_j\|^2 - d_{ij}^2 \right) \text{d}(p_i - p_j)\right). 
\end{align}
Note that 
\begin{align}
\text{d} \left(\|p_i - p_j\|^2 - d_{ij}^2 \right) = 2(p_i- p_j)^\top \text{d} p_i,\,\,\,\text{and}\,\,\,\text{d}(p_i - p_j) = \text{d}p_i. 
\end{align}
Therefore, 
\begin{align}
\text{d} \left(\|p_i - p_j\|^2 - d_{ij}^2 \right) (p_i - p_j) = 2(p_i- p_j)^\top \text{d} p_i  (p_i - p_j) = 2 (p_i - p_j) (p_i- p_j)^\top \text{d} p_i,
\end{align}
and from \eqref{eq:hessian_example1} one has
\begin{align}
\text{d}^2 V_{ij}  = (\text{d} p_i)^\top \left( 2 (p_i - p_j) (p_i- p_j)^\top  + \left(\|p_i - p_j\|^2 - d_{ij}^2 \right)\otimes I_2 \right) \text{d} p_i,
\end{align}
which readily shows the expression of Hessian matrix. We summarize:
\begin{fact}
The Hessian matrix for the potential \eqref{eq:potential_sugie1} with the gradient system \eqref{eq:gradient_example_2agent1} is identified as
\begin{align} \label{eq:Hessian_2agent}
\mathcal{H}_{{V_{ij}}_\eqref{eq:potential_sugie1}} = 2 (p_i - p_j) (p_i- p_j)^\top  + \left(\|p_i - p_j\|^2 - d_{ij}^2 \right)\otimes I_2.
\end{align}
\end{fact}


If one assumes $p_j = [0, 0]^\top$ and denotes $e_{ij} = \|p_i - p_j\|^2 - d_{ij}^2 = \|p_i\|^2 - d_{ij}^2$ and $p_i = [x_i, y_i]^\top$, then the above Hessian \eqref{eq:Hessian_2agent} is reduced to 
\begin{align} \label{eq:hessian_triangle1}
\mathcal{H}_{V_{ij}} &= 2 p_i p_i^\top  + e_{ij}\otimes I_2 \nonumber \\
& = \left[\begin{array}{cc}
2x_i^2  + e_{ij} & 2 x_i y_i \\
2x_i y_i & 2 y_i^2  + e_{ij} \\
\end{array}  \right ].
\end{align}
The Hessian formula \eqref{eq:hessian_triangle1} has been discussed in \cite{Sugie2018CDC} for stability analysis of a two-agent distance-based coordination control system. The derivation of Hessian 
\eqref{eq:hessian_triangle1} in \cite{Sugie2018CDC} is based on entry-wise identifications, which is in general not convenient   as compared with the above derivation using matrix calculus rules.

\subsection{Hessian identification for a three-agent coordination system}
As a further motivating example, we consider a three-agent coordination problem from \cite{Sugie2018CDC}, in which the potential function includes both  distance-based potentials and an area-based potential. The overall potential function is defined as
\begin{align} \label{eq:potential_sugie2}
V_{ijk} = \frac{1}{4} \left(\|p_k - p_i\|^2 - d_{ki}^2 \right)^2 + \frac{1}{4} \left(\|p_k - p_j\|^2 - d_{kj}^2 \right)^2 +\frac{1}{2} K (S - S^*)^2,
\end{align}
where $K$ is a positive scalar gain and
\begin{align}
S = - \frac{1}{2} (p_j-p_k)^{\top}J(p_i-p_j)=-\frac{1}{2} (p_j-p_k)^{\top}J(p_i-p_k)
\end{align}
with  $J = [0, 1; -1,0]$ defines the signed area of the triangle associated with three agents $(i, j, k)$. For notational convenience we denote $V_{ijk} = V_d +V_S$, with $V_d$ defined as the first two quadratic functions and $V_S$ the third quadratic function in \eqref{eq:potential_sugie2}. Note that the third quadratic function $V_S$ in \eqref{eq:potential_sugie2} with $S$ terms serves as a signed area constraint that involves positions of a three-agent group, which makes it different to the edge potential function \eqref{eq:potential_sugie1} that only involves two agents. In this example, by following the same problem setting as in \cite{Sugie2018CDC}, we again assume that agents $i$ and $j$ are fixed and stationary, and agent $k$'s dynamics are governed by a gradient descent control law
\begin{align} \label{eq:gradient_example_3agent}
\dot p_k =  & - \nabla_{p_k} V_{ijk} =  - \left(\frac{\partial V_d}{\partial p_k}+ K(S - S^*) \frac{\partial S }{\partial p_k} \right)\nonumber \\
=& (\|p_i - p_k\|^2 - d_{ik}^2)(p_i - p_k) + (\|p_j - p_k\|^2 - d_{jk}^2)(p_j - p_k) -\frac{1}{2} K(S - S^*) J (p_i - p_j),
\end{align}
where we have used the fact that 
\begin{align}
\text{d}S = -\frac{1}{2} (-\text{d}p_k)^\top J(p_i - p_j) = -\frac{1}{2}(p_i - p_j)^\top J \text{d} p_k,
\end{align}
which implies $\frac{\partial S }{\partial p_k} = \frac{1}{2} J (p_i - p_j)$. 

Now we identify the Hessian matrix $\mathcal{H}_V$, which holds $\mathcal{H}_V = \mathcal{H}_{V_d} + \mathcal{H}_{V_S}$, for the gradient flow \eqref{eq:gradient_example_3agent} associated with the potential \eqref{eq:potential_sugie2}. By following similar steps as in Section \ref{sec:motivating_examples_2agent}, one obtains 
\begin{align}
\mathcal{H}_{V_d} = &2 (p_k - p_i) (p_k- p_i)^\top  + 2 (p_k - p_j) (p_k- p_j)^\top  \nonumber \\
& + \left(\|p_k - p_i\|^2 - d_{ki}^2 \right)\otimes I_2 + \left(\|p_k - p_j\|^2 - d_{kj}^2 \right)\otimes I_2.
\end{align}

There also holds
\begin{align}
\text{d}^2 V_S &= (\text{d}p_k)^\top \text{d} \left(K(S - S^*)\frac{1}{2}J(p_i - p_j) \right)=  (\text{d}p_k)^\top \frac{1}{2}KJ(p_i - p_j) \text{d}S \nonumber \\
&=  (\text{d}p_k)^\top \left( -\frac{1}{4}KJ(p_i - p_j) (p_i - p_j)^\top J \right) \text{d}p_k, 
\end{align} 
which implies
\begin{align}
\mathcal{H}_{V_S} = -\frac{1}{4} K J(p_i - p_j) (p_i - p_j)^\top J.
\end{align}

We summarize the Hessian identification result in the following:
\begin{fact}
The Hessian matrix for the potential \eqref{eq:potential_sugie2} is identified as
\begin{align}
\mathcal{H}_{{V_{ijk}}_\eqref{eq:potential_sugie2}} = &2 (p_k - p_i) (p_k- p_i)^\top  + 2 (p_k - p_j) (p_k- p_j)^\top  \nonumber \\
& + \left(\|p_k - p_i\|^2 - d_{ki}^2 \right)\otimes I_2 + \left(\|p_k - p_j\|^2 - d_{kj}^2 \right)\otimes I_2 \nonumber \\
&  -\frac{1}{4} K J(p_i - p_j) (p_i - p_j)^\top J.
\end{align}
\end{fact}

If one assumes $d_{jk}^* = d_{ki}^* = d$ and $p_i = [-a, 0]^\top$, $p_j = [a, 0]^\top$, $p_k = [x, y]^\top$, the above Hessian formula reduces to the following
\begin{align} \label{eq:hessian_3agent_area}
\mathcal{H}_{V_{ij}} 
 = \left[\begin{array}{cc}
6x^2 +6a^2+2y^2-2d^2  & 4 x y \\
4xy  & 2x^2+2a^2+6y^2 -2d^2 +Ka^2 \\
\end{array}  \right ].
\end{align}
The Hessian formula \eqref{eq:hessian_3agent_area} has been discussed in \cite{Sugie2018CDC} for stability analysis of a three-agent formation control system with both distance and area constraints. As can be seen above, if the Hessian is calculated via the entry-wise approach, it is often tedious to get the right formula. 

The two examples presented in this section motivate the Hessian identification approach via matrix differentials and calculus rules. In the following sections, we will show how to derive general formulas for Hessian matrices for some typical potential functions in multi-agent coordination control.

\section{Hessian identification for edge-tension, distance-based potentials} \label{sec:Hessian_edge}
In this section we consider some typical  potential functions in multi-agent coordination control, which are defined as edge-tension, distance-based functions, for modeling multi-agent systems in a general undirected graph. 

Consider a local edge-tension potential in the form $V_{ij}(p_i, p_j)$ associated with edge $(i,j)$ that involves $p_i \in \mathbb{R}^d$ and $p_j \in \mathbb{R}^d$. If $(i,j) \notin \mathcal{E}$, we suppose $V_{ij} =0$. Furthermore, for the symmetry of coordination systems interacted in an undirected graph, we also assume that $V_{ij}(p_i, p_j) = V_{ji}(p_i, p_j)$. The overall potential for the whole multi-agent group is a summation of  local potentials over all edges, constructed by
\begin{align} \label{eq:potential_function_edge}
V = \frac{1}{2}\sum_{i=1}^{n} \sum_{j=1}^{n} V_{ij} (p_i, p_j)  =\frac{1}{2}\sum_{i=1}^{n} \sum_{j \in \mathcal N_i}^{n} V_{ij} (p_i, p_j).
\end{align}
The coefficient $\frac{1}{2}$ in the overall potential \eqref{eq:potential_function_edge} is due to the fact that each local potential $V_{ij}$ is counted twice in the underlying undirected graph. 

In this section we consider a general potential function as a function of inter-agent distances $\|p_i - p_j\|$, defined as
\begin{align}  \label{eq:potential_distance_function}
V_{ij} :=
\left\{
       \begin{array}{cc}
        V_{ij} (\|p_i - p_j\|),  &\text{ if } (i,j) \in \mathcal{E};  \\  
       0,  & \text{otherwise}.  \\
       \end{array}
      \right.
\end{align}
Such a distance-based potential function has found many applications in distributed multi-agent coordination control and has been one of the most popular functions in developing coordination potentials. Typical applications of the potentials \eqref{eq:potential_distance_function} and \eqref{eq:potential_function_edge} include multi-agent consensus \cite{olfati2007consensus}, distance-based formation control  
\cite{krick2009stabilisation, tyler_2009,ANDERSON2010139}, formation control laws with collision avoidance  \cite{anderson2007control, anderson2014counting}, multi-robotic navigation control \cite{de2016distributed}, multi-agent manipulability control \cite{KAWASHIMA2014}, and connectivity-preserving control \cite{ji2007distributed, zavlanos2011graph}, among others.

\subsection{Derivation of a general Hessian formula}

The control input for agent $i$ is a gradient-descent   control 
\begin{align} \label{eq:gradient_agenti}
\dot p_i = - \nabla_{p_i} V = -\sum_{j \in \mathcal N_i}^{n} \nabla_{p_i} V_{ij} (p_i, p_j). 
\end{align}

Note that 
\begin{align}
\text{d} \left( \|p_i - p_j\| \right) & = \text{d} \left(\sqrt{(p_i - p_j)^\top (p_i - p_j)}  \right) = \frac{1}{2} \left( (p_i - p_j) ^\top (p_i - p_j) \right)^{-\frac{1}{2}} \text{d} \left( (p_i - p_j) ^\top (p_i - p_j) \right) \nonumber \\
&= \frac{1}{\|p_i - p_j\|} (p_i - p_j)^\top \text{d} p_i.
\end{align}
Therefore,
\begin{align}
\frac{\partial V_{ij}(\|p_i - p_j\|) }{\partial p_i} &=\frac{\partial V_{ij}(\|p_i - p_j\|) }{\partial \|p_i - p_j\|} \frac{\partial \|p_i - p_j\| }{\partial p_i} \nonumber \\
 &= V_{ij}'  \frac{1}{\|p_i - p_j\|} (p_i - p_j),
\end{align}
where we have used the definition $V_{ij}' : = \frac{\partial V_{ij}(\|p_i - p_j\|) }{\partial \|p_i - p_j\|}$. 

Now an explicit form of the gradient-based control \eqref{eq:gradient_agenti} is derived as
\begin{align} \label{eq:control_consensus_like}
\dot p_i =  -\sum_{j \in \mathcal N_i}^{n} \nabla_{p_i} V_{ij} (p_i, p_j)  = - \sum_{j \in \mathcal N_i}^{n} V_{ij}'  \frac{1}{\|p_i - p_j\|} (p_i - p_j). 
\end{align}
The distributed coordination system in \eqref{eq:control_consensus_like} may be seen as a weighted multi-agent consensus dynamics, in which the weights, defined as $\omega_{k_{ij}} := V_{ij}'  \frac{1}{\|p_i - p_j\|}  : = V_{k}'  \frac{1}{\|z_k\|}$, are dependent on states (i.e., state-dependent weights); however, the control objective is encoded by the potential $V_{ij}$ and its gradient that encompasses many coordination tasks, while consensus or state agreement is only a special case.  
A compact form for the overall coordination system is derived as
\begin{align}
\dot p = -\nabla_{p} V  = - \left( (H^\top W H \otimes I_d \right) p,
\end{align}
where 
\begin{align} \label{eq:diagonal_matrix_W}
W = \text{diag} (\omega_1, \omega_2, \dots, \omega_m) = \text{diag} \left(  \frac{V_{1}'}{\|z_1\|}, \frac{V_{2}'}{\|z_2\|}, \dots, \frac{V_{m}'}{\|z_m\|}  \right).
\end{align}  

Following the matrix calculus rules in \eqref{eq:hessian_basic} one can show
\begin{align} \label{eq:hessian_edge}
\text{d}^2 V &= (\text{d} p)^\top  \text{d} \nabla_{p} V \nonumber \\
& = (\text{d} p)^\top   \left( (H^\top \text{d} W H \otimes I_d \right) p + (\text{d} p)^\top   \left( (H^\top W H \otimes I_d \right) \text{d} p,
\end{align}  
where
\begin{align}
\text{d} W = \text{diag}(\text{d}\omega_1, \text{d}\omega_2, \cdots, \text{d}\omega_m).
\end{align}

Recall that
\begin{align}
(H \otimes I_d ) p = z.
\end{align}
We can obtain a nice formula for the term $\left( (H^\top \text{d} W H \otimes I_d \right) p$ as follows.  
Note that 
\begin{align} \label{eq:formula_transform}
\left( \text{d} W H \otimes I_d \right) p &= \left( \text{d} W \otimes I_d \right) z \nonumber \\
& = \left( \text{diag}(\text{d}\omega_1, \text{d}\omega_2, \cdots, \text{d}\omega_m) \otimes I_d\right) \left[
\begin{array}{c}
z_1 \\
z_2 \\
\vdots \\
z_m 
\end{array} \right] = \left[
\begin{array}{c}
\text{d}\omega_1 z_1 \\
\text{d}\omega_2 z_2 \\
\vdots \\
\text{d}\omega_m z_m 
\end{array} \right]\nonumber \\
& = \left( \text{blk-diag}(z_1, z_2, \cdots, z_m) \right) \left[
\begin{array}{c}
\text{d}\omega_1 \\
\text{d}\omega_2 \\
\vdots \\
\text{d}\omega_m 
\end{array} \right].
\end{align}
Now by defining 
\begin{align}
Z : = \text{blk-diag}(z_1, z_2, \cdots, z_m) \in \mathbb{R}^{md \times m}
\end{align}
one obtains  $\left( \text{d} W H \otimes I_d \right) p = Z \text d \omega$ from \eqref{eq:formula_transform}. We then analyze the term $\text d \omega$. One can actually show
\begin{align}
\text d \omega_k = \omega_{k}'  \frac{1}{\|p_{i} - p_{j}\|} (p_{i} - p_{j})^\top \left( \text d p_{i} - \text d p_{j}\right)
\end{align}
Therefore, in a compact form, one can obtain
 \begin{align}
\text d \omega =  \Omega Z^\top (H\otimes I_d) \text{d} p,
\end{align}
where 
\begin{align}  \label{eq:diagonal_matrix_Omega}
\Omega = \text{diag} \left(\omega_{1}'  \frac{1}{\|z_1\|}, \omega_{2}'  \frac{1}{\|z_2\|}, \cdots, \omega_{m}'  \frac{1}{\|z_m\|} \right). 
\end{align}

Now \eqref{eq:hessian_edge} becomes
\begin{align} \label{eq:hessian_identi_distance}
\text{d}^2 V &= (\text{d} p)^\top  \text{d} \nabla_{p} V \nonumber \\
& = (\text{d} p)^\top   \left((H^\top \otimes I_d) Z \Omega Z^\top (H \otimes I_d) \right) \text{d} p + (\text{d} p)^\top   \left(  H^\top W H \otimes I_d \right) \text{d} p  \nonumber \\
& = (\text{d} p)^\top   \left((H^\top \otimes I_d) Z \Omega Z^\top (H \otimes I_d) + H^\top W H \otimes I_d \right) \text{d} p.  
\end{align}
Then according to the basic formula \eqref{eq:hessian_basic}, the Hessian matrix is obtained as the matrix in the middle of \eqref{eq:hessian_identi_distance}. 

In short, we now summarize the main result on Hessian identification for edge-tension, gradient-based distributed systems as follows.
\begin{fact}  \label{fact:general}
For the edge-tension distance-based potential function \eqref{eq:potential_function_edge} and the associated gradient-based multi-agent system \eqref{eq:gradient_agenti}, 
the Hessian matrix is identified as 
\begin{align} \label{eq:hessian_general_distance}
\mathcal{H}_{V_\eqref{eq:potential_function_edge}} &=     (H^\top \otimes I_d) Z \Omega Z^\top (H \otimes I_d)    +      \left(H^\top W H \otimes I_d \right), 
\end{align}
where $H$ is the incidence matrix for the underlying graph, $Z= \text{blk-diag}(z_1, z_2, \cdots, z_m)$, and the diagonal matrices $\Omega$ and $W$ are defined in \eqref{eq:diagonal_matrix_W} and \eqref{eq:diagonal_matrix_Omega}, respectively. 
\end{fact}

\begin{remark}
From the general and compact formula of Hessian matrix in \eqref{eq:hessian_general_distance}, one can also easily show 
the entry-wise expression of the Hessian. To be specific, the $(i,i)$-th block of the Hessian $\mathcal{H}_{V_\eqref{eq:potential_function_edge}}$ is expressed as
\begin{align}
\mathcal{H}_{V,ii} = \sum_{j \in \mathcal{N}_i}  \left(\Omega_{k,ij} (p_i - p_j)(p_i - p_j)^\top +w_{k,ij} I_d \right),
\end{align}
and the $(i,j)$-th block is expressed by 
\begin{align}
\mathcal{H}_{V,ij} = 
\left\{
       \begin{array}{cc}
        -\Omega_{k,ij} (p_i - p_j)(p_i - p_j)^\top -w_{k,ij} I_d,  &\text{ if } (i,j) \in \mathcal{E};  \\  
       0,  & \text{ if } (i,j) \notin \mathcal{E},   
       \end{array}
      \right.
\end{align}
where $\omega_{k,ij} =   \frac{V_{ij}'(\|p_i - p_j\|)}{\|p_i - p_j\|}$ and $\Omega_{k,ij} = \frac{\omega_{ij}'(\|p_i - p_j\|)}{\|p_i - p_j\|}$.  
\end{remark}


\subsection{Examples on Hessian identification for certain typical potential functions} \label{sec:examples_distance_potential}
In this subsection, we show several examples on Hessian identification for some commonly-used potential functions in coordination control. These potential functions have been extensively used in designing gradient-based coordination laws in  the literature; however, litter study was reported on their Hessian formulas. We will see how to use the general formula \eqref{eq:hessian_general_distance} in Fact \ref{fact:general} to derive a compact form of Hessian matrix for each potential function. 

\begin{example}
\textbf{Distance-based multi-agent formation control}, discussed in e.g., \cite{krick2009stabilisation}. The edge-tension distance-based potential is 
\begin{align} \label{eq:potential_distance_standard}
V_{ij} = \frac{1}{4} \left(\|p_i - p_j\|^2 -d_{ij}^2 \right)^2.
\end{align}
Then $\omega_k = \left(\|p_i - p_j\|^2 -d_{ij}^2 \right)$, and $W : = \text{diag}\left(\omega_1, \omega_2, \cdots, \omega_m \right)$.   It is clear that $\omega_k ' = \partial \omega_k(\|z_k\|)/ \partial \|z_k\| = 2\|z_k\|$ and therefore $\Omega = \text{diag}(2,2, \cdots, 2) = 2 I_d$. Thus, the Hessian matrix in this case is identified as
\begin{align}
\mathcal{H}_{V_{\eqref{eq:potential_distance_standard}}} &=     2(H^\top \otimes I_d) Z Z^\top (H \otimes I_d)    +      \left(H^\top W H \otimes I_d \right). 
\end{align}
In the context of distance-based formation control, the matrix $Z^\top (H \otimes I_d)$ is the \textbf{distance rigidity matrix} associated with the formation graph, denoted by $R$. The Hessian in the form 
\begin{align}
\mathcal{H}_{V_{\eqref{eq:potential_distance_standard}}} =     2 R^\top R  +      \left(H^\top W H \otimes I_d \right)
\end{align}
has been calculated (with different approaches) in e.g., \cite{tyler_2009, ANDERSON2010139, sun2015rigid} for formation systems with special shapes, e.g., 3-agent triangular shape or 4-agent rectangular shape.  
\end{example}

\begin{example}
\textbf{Formation control law with edge-based distance constraints}, discussed in e.g., \cite{anderson2007control, anderson2014counting, de2016distributed}. The edge-based potential is
\begin{align} \label{eq:potential_distance_collision}
V_{ij} = \frac{1}{2} \left(\|p_i - p_j\| -d_{ij} \right)^2.
\end{align}
Then $\partial V_{ij}/\partial \|p_i - p_j\| = \left(\|p_i - p_j\| -d_{ij} \right)$ and therefore $\omega_k = \frac{\left(\|p_i - p_j\| -d_{ij} \right)}{\|p_i - p_j\|}$. Thus, one can obtain the diagonal matrix $W$ as follows
\begin{align} \label{eq:Wmatrix_example2}
W : = \text{diag}\left(\frac{\|z_1\| - d_1}{\|z_1\|}, \frac{\|z_2\| - d_2}{\|z_2\|}, \cdots, \frac{\|z_m\| - d_m}{\|z_m\|} \right).
\end{align}
It is clear that $\omega_k ' = \partial \omega_k(\|z_k\|)/ \partial \|z_k\| = \frac{d_{k}}{\|z_{k}\|^2}$ and therefore 
   
\begin{align} \label{eq:Omega_example2}
   \Omega = \text{diag}\left(\frac{d_{1}}{\|z_{1}\|^3},\frac{d_{2}}{\|z_{2}\|^3}, \cdots, \frac{d_{m}}{\|z_{m}\|^3} \right).   
   \end{align}
   
 Thus, the Hessian matrix in this case is identified as
\begin{align}
\mathcal{H}_{V_{\eqref{eq:potential_distance_collision}}} &=     (H^\top \otimes I_d) Z \Omega Z^\top (H \otimes I_d)    +      \left(H^\top W H \otimes I_d \right), 
\end{align}
with $W$ and $\Omega$  defined in \eqref{eq:Wmatrix_example2} and \eqref{eq:Omega_example2}, respectively. 
\end{example}

\begin{example}
\textbf{Leader-follower manipulability control}, discussed in e.g.,\cite{KAWASHIMA2014}. The edge potential is a function in the form
\begin{align} \label{eq:potential_distance_manipu}
V_{ij}(p_i,p_j)=\frac{1}{2}\left(e_{ij}(\|p_i - p_j\|)\right)^2, \,\,\text{if}\, (i, j) \in \mathcal{E},
\end{align}
where $e_{ij}$ is a strictly increasing, twice differentiable function. Now we identify the Hessian matrix by following the above result in Fact \ref{fact:general}. Note that $\partial V_{ij}/\partial \|p_i - p_j\| = e_{ij} e_{ij}'$ where $e_{ij}' = \partial e_{ij}/\partial (\|p_i - p_j\|)$. Therefore $\omega_k = e_{k} e_{k}'/\|z_k\|$ and 
\begin{align}  \label{eq:matrixW_example3}
W : = \text{diag}\left(\frac{e_{1} e_{1}'}{\|z_1\|},   \frac{e_{2} e_{2}'}{\|z_2\|}, \cdots, \frac{e_{m} e_{m}'}{\|z_m\|} \right). 
\end{align}

It is clear that 
\begin{align}
\omega_k' &= \frac{\partial \omega_k(\|z_k\|)}{\partial \|z_k\|} = \frac{\partial \left(\frac{e_{k} e_{k}'}{\|z_k\|}\right)} {\partial \|z_k\|} \nonumber \\
& = \frac{\left( e_k' e_k' + e_{k} e_{k}'') \|z_k\| - e_{k} e_{k}' \right)}{\|z_k\|^2}.
\end{align}
Therefore, the entries of the diagonal matrix $\Omega$ is $\omega_k'/\|z_k\| = \frac{\left( e_k' e_k' + e_{k} e_{k}'') \|z_k\| - e_{k} e_{k}' \right)}{\|z_k\|^3}$. The Hessian matrix for the potential $V = \frac{1}{2}\sum_{i=1}^n \sum_{j=1}^n V_{ij} = \frac{1}{4}\sum_{i=1}^n \sum_{j=1}^n \left(e_{ij}(\|p_i - p_j\|)\right)^2$ is identified as 
\begin{align}
\mathcal{H}_{V_{\eqref{eq:potential_distance_manipu}}} &=     (H^\top \otimes I_d) Z \Omega Z^\top (H \otimes I_d)    +      \left(H^\top W H \otimes I_d \right), 
\end{align}
with $W$ defined in \eqref{eq:matrixW_example3} and $\Omega$  defined with diagonal entries of $\omega_k'/\|z_k\| $ as calculated above. 
\end{example}

\begin{example}
\textbf{Connectedness-preserving control in multi-agent coordination}, discussed in e.g., \cite{ji2007distributed, zavlanos2011graph}. The edge-based potential function takes the following form
\begin{align} \label{eq:potential_connected}
V_{ij}(p_i, p_j) = \frac{\|p_i - p_j\|^2}{\delta - \|p_i - p_j\|},
\end{align}
where $\delta$ is a positive parameter. Note that 
\begin{align}
V_{ij}' = \frac{\partial V_{ij}}{\partial \|p_i - p_j\|} = \frac{\left(2\delta - \|p_i - p_j\| \right) \|p_i - p_j\|}{(\delta - \|p_i - p_j\|)^2},
\end{align}
and therefore $\omega_k = V_{ij}'/\|z_k\| = \frac{2\delta - \|z_k\|}{(\delta - \|z_k\|)^2}$ where $z_k = p_i - p_j$. The diagonal matrix $W$ is obtained as
\begin{align} \label{eq:matrixW_example4}
W = \text{diag}\left( \frac{2\delta - \|z_1\|}{(\delta - \|z_1\|)^2},  \frac{2\delta - \|z_2\|}{(\delta - \|z_2\|)^2}, \cdots,  \frac{2\delta - \|z_m\|}{(\delta - \|z_m\|)^2}\right).
\end{align}
Furthermore, one can show
\begin{align}
\omega_k' = \frac{\partial \omega_k}{\partial \|z_k\|} = \frac{-\|z_k\|+ 3\delta}{(\delta - \|z_k\|)^3}.
\end{align}
Therefore, the diagonal matrix $\Omega$ can be obtained as
\begin{align} \label{eq:matrix_Omega_example4}
\Omega = \text{diag} \left(\frac{3\delta -\|z_1\|}{\|z_1\|(\delta - \|z_1\|)^3},  \frac{3\delta -\|z_2\|}{\|z_2\|(\delta - \|z_2\|)^3}, \cdots, \frac{3\delta -\|z_m\|}{\|z_m\|(\delta - \|z_m\|)^3} \right).
\end{align}
The Hessian matrix for the overall potential $V = \frac{1}{2} \sum_i^n \sum_j^n V_{ij}$ is identified as 
\begin{align}
\mathcal{H}_{V_{\eqref{eq:potential_connected}}} &=     (H^\top \otimes I_d) Z \Omega Z^\top (H \otimes I_d)    +      \left(H^\top W H \otimes I_d \right), 
\end{align}
with $W$ and $\Omega$  defined in \eqref{eq:matrixW_example4} and  \eqref{eq:matrix_Omega_example4}, respectively.
\end{example}

\subsection{A further example of Hessian identification for edge-based coordination potentials}
In this subsection, as a further example, we will show an alternative approach for identifying Hessian formula for an edge-based coordination potential. 

Consider the overall potential function
\begin{align} \label{eq:potential_distance_z4}
V = \sum_{k = 1}^{m} \frac{(\|z_k\|^2 - d_k^2)^2}{\|z_k\|^2},
\end{align}
where $z_k = p_i - p_j$ is the relative position vector for edge $k$. The potential function \eqref{eq:potential_distance_z4} has been discussed in e.g., \cite{tian2013global} for multi-agent formation and  coordination control with collision avoidance 

Define $\rho_k = \frac{\|z_k\|^4 - d_k^4}{\|z_k\|^4}$ and the gradient function for agent $i$ is obtained as 
\begin{align}
\nabla_{p_i} V = \sum_{j \in \mathcal{N}_i} 2 \rho_{k_{ij}} z_{k_{ij}}.
\end{align}
We may use $\rho_{k_{ij}}$ and $\rho_{k}$, $z_{k}$ and $z_{k_{ij}}$ interchangeably in the following text.   



Define a vector $\rho = [\rho_1, \rho_2, \cdots, \rho_m]^\top$ and a block diagonal matrix $Z= \text{blk-diag}(z_1, z_2, \cdots, z_m)$. Then one can obtain a compact form of the gradient
\begin{align}
\nabla_{p} V
&= 2   (H \otimes I_d)^\top Z \rho  \nonumber \\
&= 2 (H \otimes I_d)^\top (\text{diag}(\rho_k) \otimes I_2) (H \otimes I_d) p. 
\end{align}
where $\text{diag}(\rho_k)$ denotes a diagonal matrix with its $k$-th diagonal entry being $\rho_k$. As we will show later, the second line of the above equality will be particularly useful for later calculations.

We follow the matrix calculus rule to obtain a compact form of the Hessian. From the basic formula \eqref{eq:hessian_basic}, one can obtain
\begin{align} \label{eq:hessian}
 \text{d}^2 V(p) &=  2 (\text{d}p)^\top (H \otimes I_d)^\top (\text{d}Z) \rho +　2 (\text{d}p)^\top (H \otimes I_d)^\top Z 　\text{d}\rho.　
\end{align}

First note that
\begin{align}
(\text{d}Z) \rho = (\text{diag}(\rho_k) \otimes I_d) (H \otimes I_d) \text{d}p.
\end{align}
We then calculate the term $\text{d}\rho$.
To this end, we define $\alpha_k =\|z_k\|^2$ and $\alpha = [\alpha_1, \alpha_2, \cdots, \alpha_m]^\top$. It is obvious that
\begin{align}
 \frac{\partial \rho_k }{\partial \alpha_ k} & =  \frac{\partial \left(\frac{\|z_k\|^4 - d_k^4}{\|z_k\|^4} \right)}{\partial \|z_k\|^2}  = \frac{\partial \left(\frac{\alpha_k^2 - d_k^4}{\alpha_k^2}\right)}{\partial \alpha_k}  \nonumber \\
& = \frac{2 d_k^4}{\alpha_k^3}  = \frac{2 d_k^4}{\|z_k\|^6}.
\end{align}

Note that there holds
\begin{align}
\frac{\partial \alpha}{\partial p^\top} = 2 Z^\top (H \otimes I_d). 
\end{align}

Therefore one can obtain
\begin{align}
\text{d}\rho = 2 \text{diag}\left(\frac{2 d_k^4}{\|z_k\|^6} \right)  Z^\top (H \otimes I_d) \text{d} p.
\end{align}
From the above derivations one can further rewrite \eqref{eq:hessian} as
\begin{align}
\text{d}^2 V(p)  = &2 (\text{d}p)^\top (H \otimes I_d)^\top (\text{diag}(\rho_k) \otimes I_d) (H \otimes I_d) \text{d}p  \nonumber \\
& +　2 (\text{d}p)^\top (H \otimes I_d)^\top Z 　\left( \text{diag} \left(\frac{2 d_k^4}{\|z_k\|^6} \right) \right) 2 Z^\top (H \otimes I_d) \text{d} p　\nonumber \\
 = &(\text{d}p)^\top \left( \underbrace{2  (H \otimes I_d)^\top (\text{diag}(\rho_k) \otimes I_d) (H \otimes I_d) + 2  (H \otimes I_d)^\top Z 　\left( \text{diag}\left(\frac{4 d_k^4}{\|z_k\|^6} \right) \right) Z^\top (H \otimes I_d)}_{\text{the Hessian}} \right) \text{d} p,
\end{align}
in which a compact formula of the Hessian matrix is derived. 

In short, we summarize the above result in the following fact. 

\begin{fact}
For the potential function \eqref{eq:potential_distance_z4} in multi-agent coordination, the Hessian formula is identified as
\begin{align}
\mathcal{H}_{V_\eqref{eq:potential_distance_z4}} &= 2  (H \otimes I_d)^\top (\text{diag}(\rho_k) \otimes I_d) (H \otimes I_d) + 2  (H \otimes I_d)^\top Z 　\left( \text{diag}\left(\frac{4 d_k^4}{\|z_k\|^6} \right) \right) Z^\top (H \otimes I_d),
\end{align}
which can be equivalently written as
\begin{align}
\mathcal{H}_{V_\eqref{eq:potential_distance_z4}}= 2  (H \otimes I_d)^\top \left(\text{diag}\left( \rho_k  \otimes I_d  + \frac{4 d_k^4}{\|z_k\|^6} z_k z_k^\top  \right) \right) (H \otimes I_d).
\end{align}
\end{fact}

A brief calculation of the above Hessian matrix is shown in the appendix in  \cite{trinh2017comments}. The formula can also be calculated from the general formula in Fact  \ref{fact:general}, while the above calculation provides an alternative way to obtain a compact form of the Hessian matrix.

\section{Hessian matrix identification for composite potential functions}  \label{sec:Hessian_composite}

In this section we consider more complex potential functions that involve both distance-based functions and \textit{area-based} functions (which are thus termed composite potentials). 

These composite functions are examples of clique-based potentials (with edge-based functions being the simplest case, and triangular-area functions being the second simplest case). Since a general and compact form of   of Hessian formula for clique-based potentials is generally intractable, we will discuss in this section two examples of Hessian derivation for composite potentials with both distance and area functions, while  clique graphs specialize to 2-agent edge subgraph and 3-agent triangle subgraph.  These examples of such potential functions are taken from \cite{anderson2017formation}.
Nevertheless, the derivation of Hessian formula to be discussed in this section will be helpful in identifying Hessians for more general potentials for other clique-based graphs. 

\subsection{Identification example I: 3-agent coordination system with both distance and area functions}
 
Consider a 3-agent coordination system with the following  potential that includes two terms  incorporating both  distance and area constraints: 
\begin{align}  \label{eq:DAKindex}
V(p_1, p_2, p_3) = \sum_{(i,j) \in \{(1,2), (2,3), (1,3)\}} \frac{1}{4}(\|p_i - p_j\|^2 - d_{ij}^2)^2 + \frac{1}{2}K(S-S^*)^2,
\end{align}
where $d_{ij}$ is the desired distance between agents $i$ and $j$, and $S = - \frac{1}{2} (p_2-p_3)^{\top}J(p_1-p_2)=-\frac{1}{2} (p_2-p_3)^{\top}J(p_1-p_3)$ defines the signed area for the associated three agents. 
By denoting $V_1 = \frac{1}{4}e^\top e$, with $e = [e_1, e_2, e_3]^\top$ and $e_k = \|z_k\|^2 - d_k^2$ for $k = 1,2,3$ corresponding to the three edges, and 
$V_2  = \frac{1}{2}K(S-S^*)^2$, we rewrite $V= V_1 +V_2$. Therefore, 
the Hessian will have two parts $\mathcal{H}_{V} = \mathcal{H}_{V_1}+ \mathcal{H}_{V_2}$. 
According to Example 1 in Section \ref{sec:examples_distance_potential}, the first part of Hessian matrix $\mathcal{H}_{V_1}$ is readily identified as
\begin{align}  \label{eq:hessian_area1}
\mathcal{H}_{V_1} = 2R^\top R + (H^\top W H \otimes I_2),
\end{align}
where $R$ is the $3 \times 6$ rigidity matrix associated with the underlying undirected graph, and $W = \text{diag}\{e_1, e_2,  e_3\}$.

We now calculate the second part $\mathcal{H}_{V_2}$.
First note that $J^\top = -J$. One has
\begin{align}
\text{d} V_2 & =  K(S-S^*) \text{d} S \nonumber \\
& = - \frac{1}{2} K(S-S^*)  \left( (\text{d} p_2- \text{d} p_3)^{\top} J(p_1-p_2) +  (p_2-p_3)^{\top}J(\text{d} p_1- \text{d}p_2) \right)  \nonumber \\
& = - \frac{1}{2} K(S-S^*)  \left( (p_1-p_2)^{\top} J^{\top}  (\text{d} p_2- \text{d} p_3)  +  (p_2-p_3)^{\top}J(\text{d} p_1- \text{d}p_2) \right)  \nonumber \\
& = - \frac{1}{2} K(S-S^*)  \left( -(p_1-p_2)^{\top} J  (\text{d} p_2- \text{d} p_3)  +  (p_2-p_3)^{\top}J(\text{d} p_1- \text{d}p_2) \right)  \nonumber \\
& = - \frac{1}{2} K(S-S^*) 
\left[
\begin{array}{ccc}
(p_2-p_3)^{\top}J & (-p_1+p_3)^{\top}J & (p_1-p_2)^{\top} J
\end{array}
\right]
\left[
\begin{array}{c}
\text{d} p_1 \\
\text{d} p_2 \\
\text{d} p_3
\end{array}
\right].
\end{align}
Thus, the Jacobian matrix associated with $V_2$ can be written as 
\begin{align} \label{eq:jacobian}
A := - \frac{1}{2} K(S-S^*)
\left[
\begin{array}{ccc}
(p_2-p_3)^{\top}J & (-p_1+p_3)^{\top}J & (p_1-p_2)^{\top} J
\end{array}
\right].
\end{align}


From \eqref{eq:jacobian} one can obtain 
\begin{align} \label{eq:hessian_1}
(\text{d} A)^{\top} &= \text{d} \left( 
 -\frac{1}{2} K(S-S^*)
\left[
\begin{array}{c}
J^{\top}(p_2-p_3) \\
J^{\top}(-p_1+p_3) \\
J^{\top} (p_1-p_2) 
\end{array}
\right]\right) \nonumber \\
&= \text{d} \left(
\frac{1}{2} K(S-S^*)
\left[
\begin{array}{c}
J(p_2-p_3) \\
J(-p_1+p_3) \\
J (p_1-p_2)
\end{array}
\right]\right) \nonumber \\
&=  
\frac{1}{2} K(\text{d}S)
\left[
\begin{array}{c}
J(p_2-p_3) \\
J(-p_1+p_3) \\
J (p_1-p_2)
\end{array}
\right] + \frac{1}{2} K(S-S^*)
\left[
\begin{array}{c}
J(\text{d}p_2-\text{d}p_3) \\
J(-\text{d}p_1+\text{d}p_3) \\
J (\text{d}p_1-\text{d}p_2)
\end{array}
\right]. 
\end{align}
Note that 
\begin{align}
\text{d}S = -\frac{1}{2}\left[
\begin{array}{ccc}
(p_2-p_3)^{\top}J & (-p_1+p_3)^{\top}J & (p_1-p_2)^{\top} J
\end{array}
\right]
\left[
\begin{array}{c}
\text{d} p_1 \\
\text{d} p_2 \\
\text{d} p_3
\end{array}
\right],
\end{align}
and therefore
\begin{align}
\frac{1}{2} K(\text{d}S)
\left[
\begin{array}{c}
J(p_2-p_3) \\
J(-p_1+p_3) \\
J (p_1-p_2)
\end{array}
\right] = -\frac{1}{4} K\left[
\begin{array}{c}
J(p_2-p_3) \\
J(-p_1+p_3) \\
J (p_1-p_2)
\end{array}
\right] \left[
\begin{array}{ccc}
(p_2-p_3)^{\top}J & (-p_1+p_3)^{\top}J & (p_1-p_2)^{\top} J
\end{array}
\right]
\left[
\begin{array}{c}
\text{d} p_1 \\
\text{d} p_2 \\
\text{d} p_3
\end{array}
\right].
\end{align}

We then factorize the second term in \eqref{eq:hessian_1}:
\begin{align}
\frac{1}{2} K(S-S^*)
\left[
\begin{array}{c}
J(\text{d}p_2-\text{d}p_3) \\
J(-\text{d}p_1+\text{d}p_3) \\
J (\text{d}p_1-\text{d}p_2)
\end{array}
\right] =  
\frac{1}{2} K(S-S^*)
\left[
\begin{array}{ccc}
0 & J & -J  \\
-J & 0 & J \\
J & -J  & 0
\end{array}
\right]  
\left[
\begin{array}{c}
\text{d} p_1 \\
\text{d} p_2 \\
\text{d} p_3
\end{array}
\right].
\end{align}
Therefore, one can rewrite \eqref{eq:hessian_1} as
\begin{align}
(\text{d} A)^{\top} =
B \left[
\begin{array}{c}
\text{d} p_1 \\
\text{d} p_2 \\
\text{d} p_3
\end{array}
\right],
\end{align}
where 
\begin{small}  
\begin{align} \label{eq:hessian_area2}
B = \left( -\frac{1}{4} K\left[
\begin{array}{c}
J(p_2-p_3) \\
J(-p_1+p_3) \\
J (p_1-p_2)
\end{array}
\right] \left[
\begin{array}{ccc}
(p_2-p_3)^{\top}J & (-p_1+p_3)^{\top}J & (p_1-p_2)^{\top} J
\end{array}  \right] +
\frac{1}{2} K(S-S^*)
\left[
\begin{array}{ccc}
0 & J & -J  \\
-J & 0 & J \\
J & -J  & 0
\end{array}
\right]   \right)  
\end{align} 
\end{small}
is the Hessian matrix $\mathcal{H}_{V_2}$. We summarize the above result and calculations in the following:

\begin{fact}
For the composite potential function \eqref{eq:DAKindex}, its Hessian matrix is identified as $\mathcal{H}_{V} = \mathcal{H}_{V_1} + \mathcal{H}_{V_2}$, with $\mathcal{H}_{V_1}$ and $\mathcal{H}_{V_2}$ calculated as in \eqref{eq:hessian_area1} and \eqref{eq:hessian_area2}, respectively. 
\end{fact}

\subsection{Identification example II: 4-agent coordination system with both distance and triangular-area functions}
In this section, we consider a more complex  potential  with both distance and triangular-area functions in a 4-agent system (examples taken from \cite{anderson2017formation}). 
In this example, the overall potential is defined as 
\begin{align}\label{eq:Dindex2triangle}
 V(p_1, p_2,p_3, p_4)=  \frac{1}{4}   (e_{12}^2 + e_{23}^2 + e_{13}^2 + e_{24}^2 + e_{34}^2))  +\frac{1}{2} K \left((S_A-S^*_A)^2+(S_B-S^*_B)^2  \right),
\end{align}
where $e_{ij} = \|p_i - p_i\|^2 - d_{ij}^2$ for the five edges $(1,2), (2,3), (1,3), (2,4),  (3,4)$, and $S_A = - \frac{1}{2} (p_2-p_3)^{\top}J(p_1-p_2)=-\frac{1}{2} (p_2-p_3)^{\top}J(p_1-p_3)$ and $S_B = - \frac{1}{2} (p_3-p_4)^{\top}J(p_2-p_3)=-\frac{1}{2} (p_3-p_4)^{\top}J(p_2-p_4)$ defined as signed areas for the triangle subgraphs $(1,2,3)$ and $(2,3.4)$, respectively.


Write the performance index $V$ as a sum $V_1+V_2$, where $V_1$ contains the distance error terms $e_{ij}$ and $V_2$ contains the area error terms $S_A, S_B$. 
Again, according to Example 1 in Section \ref{sec:examples_distance_potential}, the first part of Hessian matrix $\mathcal{H}_{V_1}$ can be computed in a similar way, and is given by
\begin{equation} \label{eq:hessian_V1_composite2}
\mathcal{H}_{V_1}=2R^{\top}R+E\otimes I_2,
\end{equation}
where $R \in \mathbb{R}^{5\times 8}$ is the rigidity matrix associated with the underlying graph (the edge orientations having immaterial effect on the Hessian), and $E :=H^\top W H$ is the matrix calculated as 
\begin{equation}
E=\left[
\begin{array}{cccc}
e_{12}+e_{13}&-e_{12}&-e_{13}&0\\
-e_{12}&e_{12}+e_{23}+e_{24}&-e_{23}&-e_{24}\\
-e_{13}&-e_{23}&e_{13}+e_{23}+e_{34}&-e_{34}\\0&-e_{24}&-e_{34}&e_{24}+e_{34}\end{array}\right].  \nonumber
\end{equation}

We now identify the Hessian for the second part of potential function $V_2$. 
Note that 
\begin{align}
\text{d}V_2 =& K(S_A - S_A^*) \text{d} S_A + K(S_B-S_B^*) \text{d} S_B, 
\end{align}
in which one can show
\begin{align} \label{eq:SA_formula}
\text{d}S_A = -\frac{1}{2}\left[
\begin{array}{cccc}
(p_2-p_3)^{\top}J & (-p_1+p_3)^{\top}J & (p_1-p_2)^{\top} J & 0
\end{array}
\right]
\left[
\begin{array}{c}
\text{d} p_1 \\
\text{d} p_2 \\
\text{d} p_3 \\
\text{d} p_4 \\
\end{array}
\right],
\end{align}
and 
\begin{align} \label{eq:SB_formula}
\text{d}S_B = -\frac{1}{2}\left[
\begin{array}{cccc}
0 & (p_3-p_4)^{\top}J & (-p_2+p_4)^{\top}J & (p_2-p_3)^{\top} J
\end{array}
\right]
\left[
\begin{array}{c}
\text{d} p_1 \\
\text{d} p_2 \\
\text{d} p_3 \\
\text{d} p_4 \\
\end{array}
\right].
\end{align}

Now denote 
\begin{align}  \label{eq:Y_A}
Y_A=\left[\begin{array}{c} J(p_2-p_3)\\
J(-p_1+p_3)\\J(p_1-p_2)\\0
\end{array}\right],
\end{align}
From \eqref{eq:SA_formula} there holds $\text{d}S_A = \frac{1}{2}Y_A^\top \text{d}p$ and therefore $\text{d}\left(\frac{1}{2} K \left((S_A-S^*_A)^2 \right) \right) = (\text{d}p)^\top \left(\frac{1}{2} K\left(S_A-S^*_A\right)Y_A \right) $.
By following the Hessian matrix identification rule \eqref{eq:hessian_basic}, one has  
\begin{align} \label{eq:formula_d2YA}
\text{d}^2\left(\frac{1}{2} K \left((S_A-S^*_A)^2 \right) \right) = (\text{d}p)^\top \text{d} \left(\frac{1}{2}K(S_A-S^*_A) Y_A \right) =  (\text{d}p)^\top  \left(\frac{1}{2}K(\text{d} S_A) Y_A + \frac{1}{2} K(S_A - S_A^*) \text{d}Y_A\right).
\end{align}  
Note that $\frac{1}{2}K(\text{d} S_A) Y_A = \frac{1}{4}K(Y_A Y_A^\top) \text{d}p$, and 
\begin{align}
\text{d} Y_A = \left[\begin{array}{c} J(\text{d}p_2-\text{d}p_3)\\
J(-\text{d}p_1+\text{d}p_3)\\J(\text{d}p_1-\text{d}p_2)\\0
\end{array}\right] = \left[\begin{array}{cccc}
0&J&-J&0\\
-J&0&J&0\\
J&-J&0&0\\
0&0&0&0\end{array}\right] \left[\begin{array}{c}
\text{d} p_1 \\
\text{d} p_2 \\
\text{d} p_3 \\
\text{d} p_4 \\
\end{array}
\right].
\end{align}

Similarly, by denoting 
\begin{align} \label{eq:Y_B}
Y_B=\left[\begin{array}{c}
0\\
J(p_3-p_4)\\
J(-p_2+p_4)\\
J(p_2-p_3)
\end{array}
\right],
\end{align}
one can show 
$\text{d}S_B = \frac{1}{2}Y_B^\top \text{d}p$ and therefore $\text{d}\left(\frac{1}{2} K \left((S_B-S^*_B)^2 \right) \right) = (\text{d}p)^\top \left(\frac{1}{2} K\left(S_B-S^*_B\right)Y_B \right)$.  Similar to \eqref{eq:formula_d2YA}, there also holds
\begin{align}
\text{d}^2\left(\frac{1}{2} K \left((S_B-S^*_B)^2 \right) \right) =  (\text{d}p)^\top \text{d} \left(\frac{1}{2}K(S_B - S_B^*) Y_B \right) =  (\text{d}p)
^\top  \left(\frac{1}{2}K(\text{d} S_B) Y_B + \frac{1}{2} K(S_B - S_B^*) \text{d}Y_B\right).
\end{align}
Again, note that $\frac{1}{2}K(\text{d} S_B) Y_B = \frac{1}{4}K(Y_B Y_B^\top) \text{d}p$, and 
\begin{align}
\text{d} Y_B = \left[\begin{array}{c} J(\text{d}p_2-\text{d}p_3)\\
J(-\text{d}p_1+\text{d}p_3)\\J(\text{d}p_1-\text{d}p_2)\\0
\end{array}\right] = \left[\begin{array}{cccc}
0&0&0&0\\
0&0&J&-J\\
0&-J&0&J\\
0&J&-J&0
\end{array}\right]
 \left[\begin{array}{c}
\text{d} p_1 \\
\text{d} p_2 \\
\text{d} p_3 \\
\text{d} p_4 \\
\end{array}
\right].
\end{align}
The above calculation immediately shows the formula of the Hessian matrix. We now summarize:
\begin{fact}
The Hessian matrix associated with the potential function is identified as 
$\mathcal{H}_V = \mathcal{H}_{V_1} + \mathcal{H}_{V_2}$ with $\mathcal{H}_{V_1}$ given in  \eqref{eq:hessian_V1_composite2} and 
\begin{align}
\mathcal{H}_{V_2}=  \frac{1}{4}K\left( Y_AY_A^{\top}+2(S_A-S^*_A)
\left[\begin{array}{cccc}
0&J&-J&0\\
-J&0&J&0\\
J&-J&0&0\\
0&0&0&0\end{array}\right]    +Y_B Y_B^{\top}+2(S_B-S_B^*)\left[\begin{array}{cccc}
0&0&0&0\\
0&0&J&-J\\
0&-J&0&J\\
0&J&-J&0
\end{array}\right]
\right),
\end{align}
where  $Y_A$ and $Y_B$ are defined in \eqref{eq:Y_A} and \eqref{eq:Y_B}, respectively. 
\end{fact}

The Hessian formula was discussed and used in \cite{anderson2017formation} but details were not shown. A conventional way with entry-wise calculation will soon make the identification process intractable. We remark that, by following the two examples illustrated in this section, one can readily identify Hessians for more general composite potentials modelled in a  general undirected graph.

\section{Discussions and conclusions}  \label{sec:conclusions}
In this paper we present fast and convenient approaches for identifying Hessian matrix for several typical potentials in distributed multi-agent coordination control. We have advanced the `indirect' approach in the Hessian identification based on matrix differential and calculus rules, as opposed to the direct approach with entry-wise calculation. Many distributed coordination laws involve an overall potential as a summation of local  distance-based potentials over  all edges. For such edge-tension distance-based potentials, We derive a general formula for the Hessian matrix, with which Hessian formulas for several commonly-used coordination  potentials can be readily derived as special cases. We also analyze the case of composite potentials with both distance and  triangular-area functions, associated with a pair of three agents (as opposed to edge-tension potentials with two agents); two examples of Hessian identification for such potentials are discussed in detail. The advantage of using `indirect' matrix calculus approach shows its benefit as a fast and tractable identification process. The results in this paper can be a guidance in Hessian identification for other types of potentials in multi-agent coordination control.

\section*{Acknowledgement}
This work was supported by the Australian Research Council
under grant DP160104500. The authors would like to thank Prof. Brian D. O. Anderson for several insightful discussions on the topic of Hessian matrix identification. 

\bibliography{Hessian_calculation}

\begin{thebibliography}{10}

\bibitem{cao2013overview}
Y.~Cao, W.~Yu, W.~Ren, and G.~Chen, ``An overview of recent progress in the
  study of distributed multi-agent coordination,'' {\em IEEE Transactions on
  Industrial informatics}, vol.~9, no.~1, pp.~427--438, 2013.

\bibitem{knorn2016overview}
S.~Knorn, Z.~Chen, and R.~H. Middleton, ``Overview: Collective control of
  multiagent systems,'' {\em IEEE Transactions on Control of Network Systems},
  vol.~3, no.~4, pp.~334--347, 2016.

\bibitem{sugie2015gradient}
K.~Sakurama, S.~i.~Azuma, and T.~Sugie, ``Distributed controllers for
  multi-agent coordination via gradient-flow approach,'' {\em IEEE Transactions
  on Automatic Control}, vol.~60, pp.~1471--1485, June 2015.

\bibitem{Xudong_gradient}
X.~Chen, ``Gradient flows for organizing multi-agent system,'' in {\em Proc. of
  the 2014 American Control Conference}, pp.~5109--5114, June 2014.

\bibitem{krick2009stabilisation}
L.~Krick, M.~E. Broucke, and B.~A. Francis, ``Stabilisation of infinitesimally
  rigid formations of multi-robot networks,'' {\em International Journal of
  Control}, vol.~82, no.~3, pp.~423--439, 2009.

\bibitem{Sun2016exponential}
Z.~Sun, S.~Mou, B.~D.~O. Anderson, and M.~Cao, ``Exponential stability for
  formation control systems with generalized controllers: A unified approach,''
  {\em Systems \& Control Letters}, vol.~93, pp.~50 -- 57, 2016.

\bibitem{sun2017distributed}
Z.~Sun, M.-C. Park, B.~D.~O. Anderson, and H.-S. Ahn, ``Distributed
  stabilization control of rigid formations with prescribed orientation,'' {\em
  Automatica}, vol.~78, pp.~250--257, 2017.

\bibitem{anderson2017formation}
B.~D.~O. Anderson, Z.~Sun, T.~Sugie, S.-i. Azuma, and K.~Sakurama, ``Formation
  shape control with distance and area constraints,'' {\em IFAC Journal of
  Systems and Control}, vol.~1, pp.~2--12, 2017.

\bibitem{Xudong_SIAM_triangle}
X.~Chen, M.-A. Belabbas, and T.~Basar, ``Global stabilization of triangulated
  formations,'' {\em SIAM Journal on Control and Optimization}, vol.~55, no.~1,
  pp.~172--199, 2017.

\bibitem{KAWASHIMA2014}
H.~Kawashima and M.~Egerstedt, ``Manipulability of leader-follower networks
  with the rigid-link approximation,'' {\em Automatica}, vol.~50, no.~3,
  pp.~695 -- 706, 2014.

\bibitem{Zhao2017general}
S.~Zhao, D.~V. Dimarogonas, Z.~Sun, and D.~Bauso, ``A general approach to
  coordination control of mobile agents with motion constraints,'' {\em IEEE
  Transactions on Automatic Control}, vol.~63, pp.~1509--1516, May 2018.

\bibitem{sugie2018TCNS}
K.~Sakurama, S.~i.~Azuma, and T.~Sugie, ``Multi-agent coordination to
  high-dimensional target subspaces,'' {\em IEEE Transactions on Control of
  Network Systems}, vol.~5, pp.~345--358, March 2018.

\bibitem{sun2018cooperative}
Z.~Sun, {\em Cooperative Coordination and Formation Control for Multi-agent
  Systems}.
\newblock Springer, 2018.

\bibitem{zhang2017matrix}
X.-D. Zhang, {\em Matrix analysis and applications}.
\newblock Cambridge University Press, 2017.

\bibitem{harville1998matrix}
D.~A. Harville, {\em Matrix algebra from a statistician's perspective}.
\newblock Taylor \& Francis, 1998.

\bibitem{matrix_2005}
K.~Abadir and J.~Magnus, {\em Matrix algebra}.
\newblock Cambridge University Press, 2005.

\bibitem{mesbahi2010graph}
M.~Mesbahi and M.~Egerstedt, {\em Graph theoretic methods in multiagent
  networks}.
\newblock Princeton University Press, 2010.

\bibitem{bapat2010graphs}
R.~B. Bapat, {\em Graphs and matrices}, vol.~27.
\newblock Springer, 2010.

\bibitem{wiggins2003introduction}
S.~Wiggins, {\em Introduction to applied nonlinear dynamical systems and
  chaos}, vol.~2.
\newblock Springer Science \& Business Media, 2003.

\bibitem{absil2006stable}
P.-A. Absil and K.~Kurdyka, ``On the stable equilibrium points of gradient
  systems,'' {\em Systems \& control letters}, vol.~55, no.~7, pp.~573--577,
  2006.

\bibitem{Sugie2018CDC}
T.~Sugie, B.~D.~O. Anderson, and H.~Dong, ``On a hierarchical control strategy
  for multi-agent formation without reflection,'' in {\em Submitted to the 2018
  IEEE Conferences on Decision and Control}, 2018.

\bibitem{olfati2007consensus}
R.~Olfati-Saber, J.~A. Fax, and R.~M. Murray, ``Consensus and cooperation in
  networked multi-agent systems,'' {\em Proceedings of the IEEE}, vol.~95,
  no.~1, pp.~215--233, 2007.

\bibitem{tyler_2009}
T.~H. Summers, C.~Yu, B.~D.~O. Anderson, and S.~Dasgupta, ``Formation shape
  control: Global asymptotic stability of a four-agent formation,'' in {\em
  Proceedings of the 48h IEEE Conference on Decision and Control (CDC) held
  jointly with 2009 28th Chinese Control Conference}, pp.~3002--3007, Dec 2009.

\bibitem{ANDERSON2010139}
B.~D.~O. Anderson, C.~Yu, S.~Dasgupta, and T.~H. Summers, ``Controlling four
  agent formations,'' in {\em The 2nd IFAC Workshop on Distributed Estimation
  and Control in Networked Systems}, pp.~139 -- 144, 2010.

\bibitem{anderson2007control}
B.~D.~O. Anderson, C.~Yu, S.~Dasgupta, and A.~S. Morse, ``Control of a
  three-coleader formation in the plane,'' {\em Systems \& Control Letters},
  vol.~56, no.~9-10, pp.~573--578, 2007.

\bibitem{anderson2014counting}
B.~D.~O. Anderson and U.~Helmke, ``Counting critical formations on a line,''
  {\em SIAM Journal on Control and Optimization}, vol.~52, no.~1, pp.~219--242,
  2014.

\bibitem{de2016distributed}
H.~G. de~Marina, B.~Jayawardhana, and M.~Cao, ``Distributed rotational and
  translational maneuvering of rigid formations and their applications,'' {\em
  IEEE Transactions on Robotics}, vol.~32, no.~3, pp.~684--697, 2016.

\bibitem{ji2007distributed}
M.~Ji and M.~Egerstedt, ``Distributed coordination control of multi-agent
  systems while preserving connectedness,'' {\em IEEE Transactions on
  Robotics}, vol.~23, no.~4, pp.~693--703, 2007.

\bibitem{zavlanos2011graph}
M.~M. Zavlanos, M.~B. Egerstedt, and G.~J. Pappas, ``Graph-theoretic
  connectivity control of mobile robot networks,'' {\em Proceedings of the
  IEEE}, vol.~99, no.~9, pp.~1525--1540, 2011.

\bibitem{sun2015rigid}
Z.~Sun, U.~Helmke, and B.~D.~O. Anderson, ``Rigid formation shape control in
  general dimensions: an invariance principle and open problems,'' in {\em
  Proc. of the 2015 IEEE 54th Annual Conference on Decision and Control (CDC)},
  pp.~6095--6100, IEEE, 2015.

\bibitem{tian2013global}
Y.-P. Tian and Q.~Wang, ``Global stabilization of rigid formations in the
  plane,'' {\em Automatica}, vol.~49, no.~5, pp.~1436--1441, 2013.

\bibitem{trinh2017comments}
M.~H. Trinh, V.~H. Pham, M.-C. Park, Z.~Sun, B.~D.~O. Anderson, and H.-S. Ahn,
  ``Comments on “global stabilization of rigid formations in the plane
  [automatica 49 (2013) 1436--1441]”,'' {\em Automatica}, vol.~77,
  pp.~393--396, 2017.

\end{thebibliography}
\bibliographystyle{ieeetr}

\end{document}